# Crack propagation threshold in single-crystal silicon: a cleavage plane-crackons model

Faming Gao

State Key Laboratory of Bio-based Fiber Materials, Tianjin University of Science and Technology,

Tianjin 300222, China

**Abstract:** Experiments revealed the discontinuities (speed gap) in the dependence of the crack speeds on the crack driving force. Despite great efforts, until now, previous theoretical methods, such as linear elastic fracture mechanics and molecular dynamics calculations, failed to elucidate the issue of speed gap. Herein, the cleavage plane-crackons model for crack propagation has been proposed. The normalized crack propagation speed of silicon is proportional to the one-fourth power of the quantum numbers *N*: $v/c_R = (2N/\pi^3)^{1/4}$. This motion equation is consistent with experimental results of silicon. It clarifies the underlying mechanism of the long-standing issue of the "speed gap". The relationship between quantum numbers and the roughness of the surface of the cracks has been established. The critical quantum number for the onset of unstable crack propagation in single-crystal silicon has been determined. The stress intensity factor *K* can be expressed as, $K = c_R\sqrt{2N\rho\gamma_0}$. The correspondence between the critical quantum number, $N_{\text{crit}} = 5$, and fracture toughness, $K_c = c_R\sqrt{10\rho\gamma_0}$, has been discovered. This methodology lead to insights into the underlying mechanism of the fracture processes. It is not limited to silicon and can be extended to other crystalline material to understand and predict the fracture behaviors.

**Contact author:** fmgao@tust.edu.cn

Most fracture study is carried out in the elegant framework of linear elastic fracture mechanics (LEFM) [1-5]. In particular, using LEFM the equation of motion for the straight, dynamic crack is derived [6],

$$\frac{G_d}{G_s} = 1 - \frac{v}{c_R} \tag{1}$$

where $G_s$ is the static strain energy release rate and $c_R$ is the Rayleigh wave speed in the direction of crack propagation. In the simplest case, the dynamic fracture energy $G_d$ is a constant, $G_c$, equal to the specific surface energy $2\gamma_0$ of the two fracture surfaces created by the advancing crack. According to Eq. (1), the speed, $v$ of a dynamic crack is a monotonic, continuous function of the static strain energy release rate $G_s$ [6]. In contrast, the threshold of crack speed exists. Hauch et al [7] perform the meticulous fracture experiment of single crystal silicon, and results indicated that under the critical load cracks jump to a steady state crack speed of 2 km/s. This speed gap demonstrates that crack speeds between zero and 2 km/s are unallowed. To explain these issues, molecular dynamics calculations (MD), a claimed powerful investigative tool, are frequently employed to simulate fracture properties [8-13]. However, all of the calculated MD speed gaps are substantially larger than are one observed in experiments. Bernstein et. al. used quantum mechanics method to get the correct energy. But, at higher loads their results fail to show an increase of speed with load, this is not in agreement with experimental results [7]. Nevertheless, it is imply the speed gaps of crack system could be a quantum mechanics effect, the quantization for crack propagation seems to be a feasible method.

Crystal cleavage refers to the property of a crystal to split along specific crystallographic directions under the action of external force, forming smooth planes. These planes of fracture are called cleavage planes. Orowan [14] assumed that elastic deformation is constrained by two adjacent cleavage planes and that the material responds linearly to external applied load (Hooke's behavior). He pointed out that once the interatomic spacing reaches a critical value, all atoms separate simultaneously. Normally, the chemical bond between two atoms are one-dimensional linear bond. In order to study the vibrations of such one-dimensional linear bonds, One simply analogize the chemical bond to a spring connecting two small balls (atoms) and understand the vibrational behavior of the bond through the spring-based harmonic oscillator model. According to Orowan's hypothesis, we can regard the cleavage planes system as a giant spring connecting two cleavage planes, where the two cleavage planes are equivalent to the two small balls (atoms) of a one-dimensional linear bond. Furtherly, we can assume that the chemical bond between the cleavage planes in materials is regarded as a cleavage plane-bond. The vibration of the cleavage plane-bond is simplified into the ideal case of a quantum harmonic oscillator. The spacing between planes and the specific surface energy are the characteristic quantities of cleavage plane-bond. For example, the cleavage planes of silicon single crystal are {111} and {110} planes, which correspond to {111} cleavage plane-bond and {110} cleavage plane-bond, respectively. If the potential energy of the cleavage plane-bond is $V(r) = \frac{1}{2}k(r - r_e)^2$, where $r_e$ is the equilibrium distance of planes, we can employ a harmonic oscillator to describe vibrational behavior of the cleavage plane-bond. Hamiltonian of harmonic oscillator is $H = -\frac{\hbar^2}{2M}\frac{\partial^2}{\partial r} + \frac{1}{2}k(r - r_e)^2$, where $M$ is the reduced mass of the cleavage plane-bond. By solving Schrödinger's equation, $H\Psi(r) = E\Psi(r)$, we can get the energy eigenvalues $E_n$ of the cleavage plane-bonds,

$$E_n = (n + \tfrac{1}{2})\hbar\sqrt{k/M} \tag{2}$$

where the quantum number $n$ = 0, 1, 2, 3, ···. $\hbar\sqrt{k/M}$ is the characteristic energy of the cleavage plane-bond. When the plane-bond absorb the energy, the bonding state will transition to a higher energy level. When the plane-bond absorb the enough energy, the transition of states would result in the breaking of plane-bond. In this bond breaking case, a crack could be generated along the cleavage plane, and the new surface will be formed. According to Griffith's fracture theory, the crack propagation would create two new surface. The energy required to form a new surface is called as the surface energy $\gamma_0$. The crack driving force, Griffith's strain energy release rate $G_c$ is expressed as $G_c = 2\gamma_0$. It is noted that Holland et. al. [12] suggested a term of "crackon" to describe the crack behavior. Herein, we define a cleavage plane-crackon to describe the quantized behavior of cleavage fracture. The specific surface energy is its characteristic energy. If we assume that the $E_0$ has the following relationship with specific surface energy: $E_0 = \frac{1}{2}\hbar\sqrt{k/M} = A(n)\gamma_0$, or $k = 2M(\frac{A(n)\gamma_0}{\hbar})^2$, the energy eigenvalues of the cleavage plane-crackon can be expressed as,

$$E_N = 2A(n)\gamma_0\left(N + \tfrac{1}{2}\right), N = 0, 1, 2, 3, \cdots \tag{3}$$

According to Eq. (3), when the cleavage plane-crackon of silicon absorb the energy of $\Delta E_N = E_N - E_0 = 2NA(n)\gamma_0$, the cracks will be generated and propagate to balance these energy. Griffith's strain energy release rate, $G_c = 2\gamma_0 = (E_1 - E_0)/A(n)$. The static strain energy release rate $G_s$ of silicon can be calculated by the transition energy between the energy levels of the cleavage plane-crackon as follow,

$$G_s = (E_N - E_0)/A(n) = 2N\gamma_0 \tag{4}$$

In the cohesive strength model [15, 16], the surface energy is expressed as

$$\gamma_0 = \frac{E' d_0}{\pi^2} \tag{5}$$

where $E'$ is the appropriate elastic modulus, $d_0$ is the effective bond spacing. On substituting Eq.(4) and Eq. (5) into the fracture equation, that is

$$\sqrt{G_s E'} = \sqrt{\frac{2N}{\pi^3}}E'\sqrt{\pi d_0} = \sigma\sqrt{\pi d_0} \tag{6}$$

Where

$$\sigma = \sqrt{\frac{2N}{\pi^3}}E' \tag{7}$$

According to Ref. [17, 18], fracture speed can be expressed as

$$v = \sqrt{\frac{\sigma}{\rho}} \tag{8}$$

Thus, Eq. (6) is rewritten as

$$\sqrt{G_s E'} = \rho v^2 \sqrt{\pi d_0} \tag{9}$$

On substituting Eq.(7) into Eq. (8),

$$v = \sqrt[4]{\frac{2NE'^2}{\pi^3 \rho^2}} \tag{10}$$

For transition of a cleavage plane-crackon from the ground state *N*=0 to state *N*=1, $G_s = G_c$. Then

$$\sqrt{G_c E'} = \sqrt{\frac{2}{\pi^3}}E'\sqrt{\pi d_0} \tag{11}$$

By comparing Eq. (9) and Eq. (11), we can derive the equation of the crack motion:

$$v\sqrt{\frac{\rho}{E'}} = \sqrt[4]{\frac{2}{\pi^3}\frac{G_s}{G_c}} \tag{12}$$

For a long time，Rayleigh wave speed $c_R$ is regards as the upper limit of crack speed in crystals. Rayleigh waves are interface elastic waves formed by the coupling of longitudinal and transverse (shear) waves, propagating along the free surface of an elastic half-space. The longitudinal and shear wave speeds is $v_L = \sqrt{(B+4G/3)/\rho}$ and $v_s = \sqrt{G/\rho}$, respectively, where $\rho$ is density, *B* is bulk modulus, *G* is shear modulus [19]. Thus, $c_R$ depends on the modulus *B* and *G,* and may express as: $c_R = \sqrt{f'(G,B)/\rho}$ like $v_L$. In Ref. [16] the appropriate elastic modulus $E'$ is suggested as $E' = f(G,B)$. From the experimental elastic constants ($C_{11}$=165.7 GPa, $C_{12}$=63.9 GPa, $C_{33}$=79.56 GPa) [20], the appropriate elastic modulus for silicon (111) plane can be determined, $E' = 51.4$ GPa. The experimental value of $c_R$ is 4.5-4.7 km/s [8-10]. Then, $\sqrt{E'/\rho} =$ 4.698 km/s $\approx c_R$. Thus, the equation of the crack motion of silicon, Eq. (12) is expressed as

$$\frac{v}{c_R} \approx \sqrt[4]{\frac{2}{\pi^3}\frac{G_S}{G_C}} \tag{13}$$

Distinctly, Eq. (13) is different from Eq. (1) established by LEFM. Fracture speeds $v$ as the function of an applied load $G_s$ from Eq. (13) appear in Fig. 1 along with other calculation and experimental results [7]. From Fig. 1, it can be seen that MD calculations produce substantially larger velocity gaps, about 5.0 J/m$^2$ than that of experiment, 2.0-2.5 J/m$^2$ [7, 13]. Although the tight-binding quantum mechanics calculation by Bernstein and Hess [11] get the 85% onset energy of fracture, but it has failed to produce a prediction for the experimental results at high speeds. Our computational results (full square ) by Eq. (13) are in agreement with experimental results of fracture speeds. At the critical load $G_s = 2\gamma_0$, our calculated crack speed is 2.3 km/s, which is in agreement with the experimental values, 2.0 km/s±0.2 km/s [7].

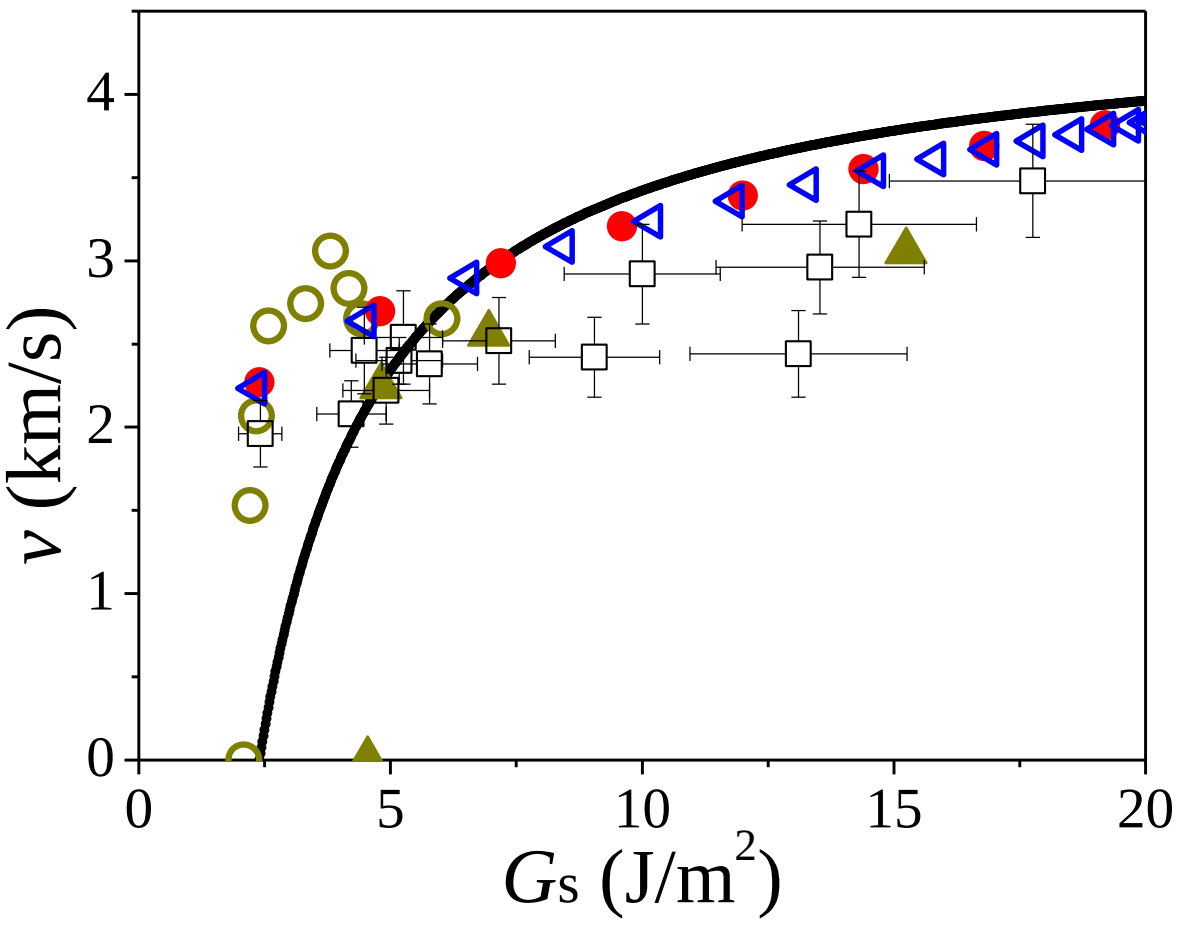

FIG. 1. Crack speed ($v$) as a function of static energy release rate ($G_s$) and Ref. [10] (full triangle), Ref. [11] (open circle), this work by harmonic oscillator potential (full circle) and this work by Morse potential (open triangle), compared with the experimental results (open square) of Hauch et al. [7]. Solid line is prediction from LEFM [Eq. (1)].

Since anharmonicity exists in actual vibrations, it is generally believed that replacing the harmonic potential function with the Morse potential can better reflect the actual situation of vibrations. The Morse potential is [21, 22],

$$V(r) = De^{-2a(r-r_0)} - 2De^{-a(r-r_0)} \tag{14}$$

where $D$ and $a$ are parameters. The energy levers of the cleavage plane-crackon is as follows,

$$E_n = -D + 2A(n)\gamma_0[(n + \tfrac{1}{2}) - \tfrac{1}{\eta_e}(n + \tfrac{1}{2})^2] \quad n = 0, 1, 2, 3, \cdots \tag{15}$$

The quantum number $n$ has an upper limit, $n < \frac{1}{2}(\eta_e - 1)$. The static strain energy release rate $G_s$ of silicon can be expressed as follow,

$$G_s = 2\gamma_0 n[1 - \tfrac{1}{\eta_e}(n + 1)] = 2N\gamma_0 \tag{16}$$

$$N = n[1 - \tfrac{1}{\eta_e}(n + 1)] \tag{17}$$

Fracture speeds $v$ as the function of an applied load $G_s$ calculated from Eqs. (16-17) appear in Fig. 1. From Fig. 1, it can be seen that the calculated results is very similar using the Morse potential and the harmonic oscillator potential. Therefore, we could still use the simplicity of harmonic oscillators to understand cleavage fracture behavior.

The continuum mechanics theory assumes that speed changes monotonically with load. However, Fig. 1 shows that the measured speeds appear a certain dispersivity at higher load points, and some variation even starts from the lowest speed of 2 km/s. The continuum mechanics theory does not provide an explanation for this. According to the present study, when the load is $G_s = G_c$, the corresponding quantum number $N$ transitions from the ground state $N = 0$ to state $N = 1$. In this case, there is only one possible transition, so the measured speed is monotonic. When $G_s$ reaches $2G_c$, there are two possible transitions: one is from the state $N = 0$ to state $N = 2$, where the speed is about 2.6 km/s; the other is a transition from the state $N = 0$ to state $N = 1$, where the speed is about 2.3 km/s. Similarly, when $G_s = NG_c$, there are N possible transitions. Therefore, due to probable quantum transition, in experiments, the measured speeds could show a wide distribution. At the lowest load, point $G_c$, the measured speed appears the only speed value, showing a clear monotonic relationship.

The crack instabilities cannot be understood either by linear elastic fracture mechanics. Cracks moving of cleavage plane at low speeds may create atomically mirrorlike surfaces. At higher speeds, rough, irregularly surfaces can be formed. To understand the underlying physical mechanism, the dynamical instability of brittle materials are studied theoretical and experimentally [23-28]. Cramer et al [8] provided a relation between the crack dynamics and the fracture surface morphology of single crystalline silicon. At the lowest fracture stress $G_s$ = 7 J/m$^2$, the fracture surface of (110) cleavage plane is mirrorlike. At above crack speeds of $v$ = 0.67$c_R$ the fracture surface appears hills and valleys. To study the morphologies of the fracture surface, Wang et al [28] carried out detail experiment on the (110) cleavage plane of silicon samples at different speeds. Their work (see Fig. 1D in Ref. [28]) shows that the fracture surface exhibit the mirrorlike morphology at $v \leq 0.59c_R$ and the presence of corrugations a $v \geq 0.67c_R$ for the surface-polished samples.

Table I. Morphologies of the fracture surface of silicon (110) plane, static strain energy release rates $G_s$(J/m$^2$), fracture speeds $v$(km/s), and quantum numbers *N*.

| N | $v$ | $G_s$ | Morphology ($G_{s\,\mathrm{exp}}$) [8] | Morphology ($v_{\mathrm{exp}}$) [28] |
|---|---|---|---|---|
| 1 | 0.504$c_R$ | 3.4 | | mirrorlike (0.50$c_R$) |
| 2 | 0.599$c_R$ | 6.8 | mirrorlike (7) | mirrorlike (0.59$c_R$) |
| 3 | 0.663$c_R$ | 10.2 | hills, valleys | corrugations (0.67$c_R$) |
| 4 | 0.713$c_R$ | 13.6 | hills, valleys (14) | |
| 5 | 0.754$c_R$ | 17 | hills, valleys (17.8) | corrugations (0.74$c_R$) |
| 6 | 0.789$c_R$ | 20.4 | | jagged patterns, instability (0.80$c_R$) |
| 7 | 0.820$c_R$ | 23.8 | | |
| 8 | 0.848$c_R$ | 27.2 | hills, valleys (26.8) | |

In Table I, we compare the static strain energy release rates $G_s$ and fracture speeds $v$ calculated by Eq. (13) with experimental results [8, 28]. In Ref. [8], $G_c$= 3.4 J/m$^2$. At quantum number $N \leq 2$, fracture speed $v \leq 0.599c_R$ or static strain energy release rate $G_s \leq$ 6.8 J/m$^2$, the fracture surface of the (110) cleavage plane exhibit the mirrorlike morphology. At quantum number $N \geq 2$, or $v \geq 0.663c_R$ or $G_s \geq$ 10.2 J/m$^2$, the fracture surface appears corrugations.

Wang et. al. [28] pointed out that dynamic crack instabilities occur at $v > 0.74c_R$ or $v = 0.8c_R$, and the jagged patterns appears on the fracture surface at $v = 0.8c_R$. In addition, Yoffe's model [29] of the moving crack predicts the instability onset occurs at about 73% of the Rayleigh-wave speed. Therefore, it could be an acceptable suggestion that the dynamic crack instabilities of the (110) cleavage plane of silicon occur at quantum number $N > 5$, or $v > 0.754c_R$ or $G_s > 17$ J/m$^2$. Beyond a critical crack speed, $v_{crit} \approx 0.754c_R$, or a critical quantum number $N_{crit} = 5$, an intrinsic instability would be observed.

The stress intensity factor $K$ can be expressed as

$$K = \sqrt{G_s E'} = c_R\sqrt{\rho N G_c} \tag{18}$$

We can calculate the stress intensity factors $K$ on (111) Si using Eq. (18). The $c_R$ have been measured on (111) surfaces of silicon single crystal [30, 31]. Fig. 2 shows the dependence of the stress intensity factors on the propagation direction, which was plotted with respect to the [110] direction ( $0^{o}$ ). The $30^{o}$ corresponds to propagation along the [112] direction. The threshold value of the stress intensity factor $K$ is the toughness $K$c, which are material constant, to be measured experimentally. Chen et. al. measured fracture toughness ( $K$c ) for the (111) [11-2] cleavage systems of silicon crystal, $K$c = 0.82 MPa m$^{1/2}$ [32]. From Fig. 2, it can be seen that the measured fracture toughness $K$c corresponds to the stress intensity factor $K$ at the critical quantum number, $N_{crit} = 5$.

It implies that we may employ the fracture toughness $Kc$ to characterize the intrinsic instability. As quantum number $N = 5$, Eq. (4) can be written as,

$$\sqrt{5G_cE'} = \sqrt{\frac{10}{\pi^3}}E'\sqrt{\pi d_0} = \sigma\sqrt{\pi d_0} \tag{19}$$

It is worth noting that Eq. (19) can be transfer as Eq. (14) in Ref. [16] due to $\pi^2 \approx 10$,

$$\sqrt{\frac{\pi^2}{2}G_cE'} = \frac{1}{\sqrt{\pi}}E'\sqrt{\pi d_0} = \sigma\sqrt{\pi d_0} \tag{20}$$

$$K_c = \sqrt{\frac{\pi^2}{2}G_cE'} = \frac{1}{\sqrt{\pi}}E'\sqrt{\pi d_0} \tag{21}$$

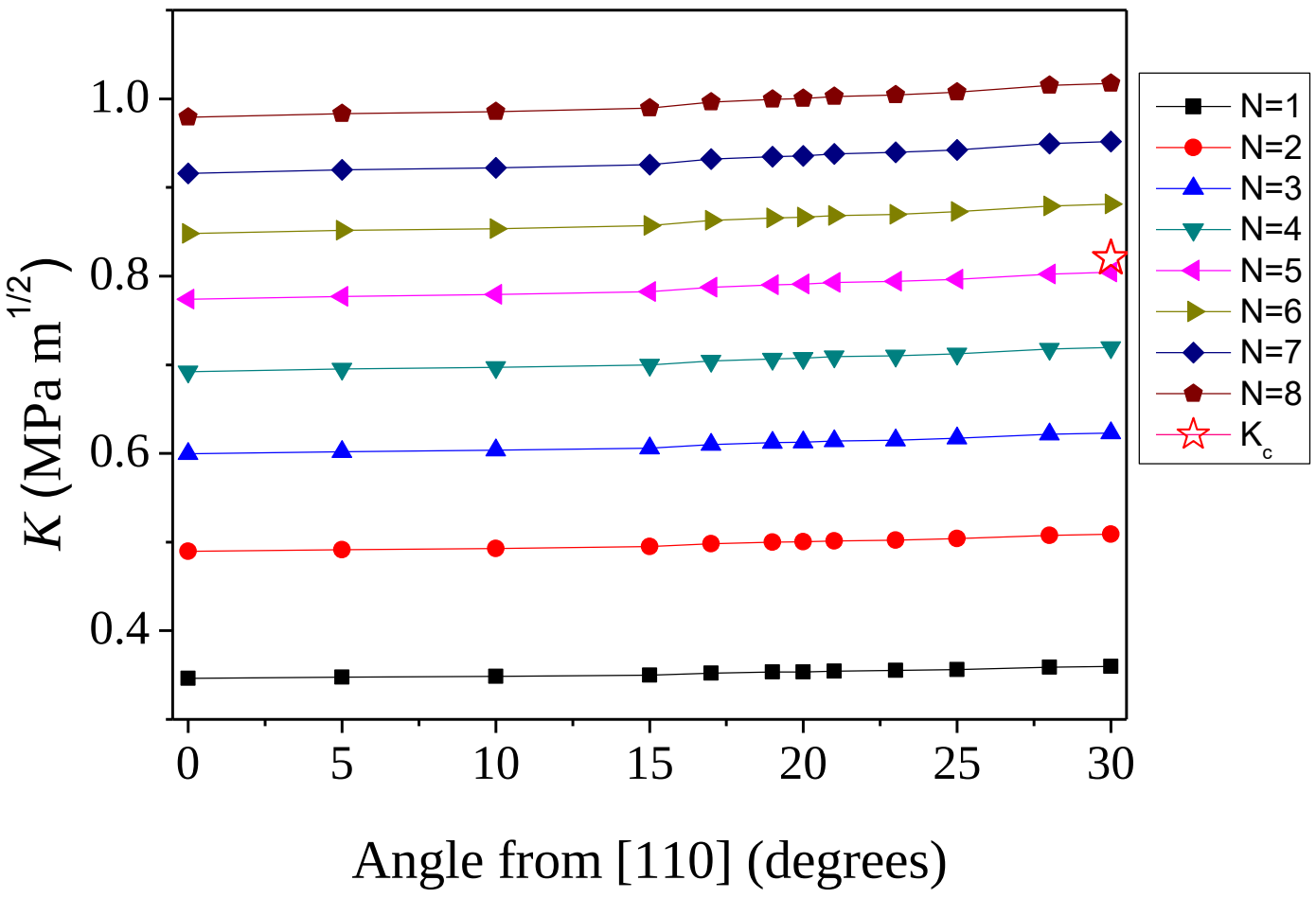

FIG. 2. Angular dependence of stress-intensity fractors ($K$, MPa m$^{1/2}$) on (111) Si, for several quantum number ($N = 1\sim8$). Open pentagram is the experimental fracture toughness ( $K_c$ ) for the (111) [11-2] cleavage systems of silicon crystal [32]. The solid lines are only a guide for the eye.

In Ref. [16], the Eqs. (20-21) has been used to calculate the fracture toughness of 28 typical ionic and covalent single crystals, the calculated results are in agreement with the measured values [20]. That is to say, Eq. (19) in this work can be employed to calculate the fracture toughness of 28 typical ionic and covalent single crystals. These demonstrate that our methodology here is not limited to silicon and could be extended to other crystalline material to understand and predict the fracture properties.

In summary, a harmonic oscillator model for crack propagation has been proposed. The static strain energy release rate is defined as the transition energy of the cleavage plane-crackons. The energy levels of cleavage plane-crackons in silicon is derived via solving the hamiltonian equation of harmonic oscillator. Based on the quantum transition of crackons the new equation of motion for cleavage crack is established, $v/c_R = (2N/\pi^3)^{1/4}$, which is capable of elucidating experimental observations including crack speed gap, fracture morphology and instability. This work has opened a new avenue for exploring the quantization fracture mechanisms.

**ACKNOWLEDGMENTS**

The author acknowledges the financial support from the National Natural Science Foundation of China (Grant No. 22379112, 21071122)

**AUTHORDECLARATIONS**

**Conflict of Interest**

The authors have no conflicts to disclose.

**REFERENCE**

[1] A. A. Griffith, Phil. Trans. Roy. Soc. Lond. A, 221, 163 (1921).

[2] G. R. Irwin, Trans. ASME, Ser. E, J. Appl. Mech. 24, 361 (1957).

[3] E. Orowan, Rep. Prog. Phys., 12, 185 (1949).

[4] L. I. Slepyan, Sov. Phys. Dokl. 26, 538(1981)..

[5] R. Thomson, Solid State Physics 39, 1 (1986).

[6] L.B. Freund, Dynamic Fracture Mechanics (Cambridge University Press, Cambridge, England,

1990).

[7] J.A. Hauch, D. Holland, M.P. Marder, H. L. Swinney, Phys. Rev. Lett. 82, 3823 (1999).

[8] T. Cramer, A. Wanner, P. Gumbsch, Phys. Rev. Lett. 85, 788 (2000).

[9] J.G. Swadener, M.I. Baskes, M. Nastasi, Phys. Rev. Lett. 89, 085503 (2002)

[10] M. J. Buehler, H. Tang, A. C. Van Duin, W. A. Goddard III, Phys. Rev. Lett. 99, 165502 (2007)

[11] N. Bernstein, D. W. Hess, Phys. Rev. Lett. 91, 025501(2003).

[12] D. Holland, M. Marder, Adv. Mater. 11, 793(1999).

[13] E. Bouchbinder, J. Fineberg, M. Marder, Annu. Rev. Condens. Matter Phys. 1, 371(2010).

[14] E. Orowan, Fracture and strength of solids, Rep. Prog. Phys., 12, 185 (1949).

[15] M. A. Meyers, K.K. Chawla, Mechanical behavior of materials (Cambridge University Press, New York, 2008), 2nd ed.

[16] F. M. Gao, Phys. Status Solidi (RRL), 20, e70216 (2026).

[17] H. Gao, J. Mech. Phys. Solids 44, 1453 (1996).

[18] M.J. Buehler, H. Gao, Nature 439, 307(2006).

[19] T. Goto, O. L. Anderson, J. Geophys. Research, 94, 7588 (1989)

[20] W. H. Harrison, Electronic structure and the properties of solids (Freeman, San Francisco, 1980).

[21] P. M. Morse, Phys. Rev. 34, 57(1929).

[22] Q. Gu, Quantum Mechanics I (Science Press, Beijing, 2014)

[23] L. Zhao, D. Bardel, A. Maynadier, D. Nélias, Scripta Mater. 130, 83 (2017).

[24] D. Bardel, A. Maynadier, D. Nélias, Nat. Commun. 9, 1298 (2018).

[25] A. Gleizer, G. Peralta, J. R. Kermode, A. De Vita, D. Sherman, Phys. Rev. Lett. 112,

115501(2014).

[26] M. Wang, L. Zhao, M. Fourmeau, D. Nelias, Z. H. Li, Int. J. Fract. 245, 157(2024)

[27] F. Atrash, A. Hashibon, P. Gumbsch, D. Sherman, Phys. Rev. Lett. 106, 085502 (2011).

[28] M. Wang, M. Fourmeau, L. Zhao, F. Legrand, and D. Nélias, Proc. Natl. Acad. Sci. USA 117, 16872 (2020).

[29] E. H. Yoffe, Phil. Mag. 42, 739 (1951).

[30] R. G. Pratt and T. C. Lim, Appl. Phys. Lett. 15, 403 (1969)

[31] M.H. Kuok, S.C. Ng, Z.L. Rang, T. Liu, Solid State Commun. 110, 185 (1999)

[32] C.P. Chen, M.H. Leipold, Bull. Am. Cer. Soc. 59, 469 (1980).